\documentclass{iau} 
\usepackage{graphicx}

\title[IAUS.\  317.~~Globular Clusters and Dwarf Galaxy Nuclei] 
{Are the globular clusters with significant internal [Fe/H] spreads all former dwarf galaxy nuclei?}

\author[G. S. Da Costa]   
{G. S. Da Costa$^1$}

\affiliation{$^1$Research School of Astronomy \& Astrophysics, Australian National University \\ email: {\tt gary.dacosta@anu.edu.au}} 

\pubyear{2015}
\volume{317}  
\jname{The General Assembly of Galaxy Halos: Structure, Origin and Evolution}
\editors{A. Bragaglia, M. Arnaboldi, M. Rejkuba \& D. Romano, eds.}
\begin{document}

\maketitle

\begin{abstract}
In this contribution the hypothesis that the Galactic globular clusters with substantial internal [Fe/H] abundance
ranges are the former nuclei of disrupted dwarf galaxies is discussed.  Evidence considered includes the form of the
metallicity distribution function, the occurrence of large diffuse outer envelopes in cluster
density profiles, and the presence of \mbox{([$s$-process/Fe], [Fe/H])} correlations.  The hypothesis is shown to be 
{\it plausible} but with the caveat that if significantly more than the current nine clusters known to have [Fe/H] 
spreads are found, then re-evaluation will be required.
\end{abstract}

\firstsection 
\section{Introduction}

The current standard picture of the formation of our Galaxy, and particularly its stellar halo, postulates that there is a 
substantial contribution from the disruptive merger and accretion of dwarf galaxies.  Dwarf galaxies often possess
their own globular cluster systems, and in some cases they also have nuclei/nuclear star clusters 
(\cite[e.g., den Brok et al.\ 2014]{denBrok14}).  Consequently, these dwarf galaxy globular and nuclear star clusters 
will be accreted into the halo of the Galaxy as the dwarf is tidally disrupted. 
The on-going disruption of the Sagittarius  (Sgr) dwarf by the Milky Way is an example of this process in action.  
There are a number of globular clusters currently associated with this dwarf 
(\cite[e.g., Law \& Majewski 2010]{LM10}), and the cluster M54 lies at the dwarf galaxy's centre.  When the Sgr 
accretion process is complete, these clusters will simply become part of the 
Galaxy's halo globular cluster population and their origin in Sgr will not be easily established.

The vast majority of Galactic globular clusters have one characteristic in common, and that is that their constituent
stars are chemically homgeneous as regards the abundances of the heavier elements such as Fe and Ca.  There 
are, however, a small number of ``globular clusters'' that show definite intrinsic internal [Fe/H] abundance 
dispersions.  In this contribution the hypothesis that these systems are the former nuclear star clusters of now
disrupted dwarf galaxies is investigated (see also \cite[Willman \& Strader 2012]{WS12} and 
\cite[Marino et al. 2015]{AM15}).
Before embarking on this task, however, there are a number of caveats that need to be mentioned.  
First, \cite[Yong et al.\ (2013)]{DY13} have used an exquisite level of analysis to show that in the 
globular cluster NGC~6752, there are real star-to-star [Fe/H] abundance variations at very low levels ($\Delta$[Fe/H]
$\sim$ 0.03 dex).  It is possible that all clusters show [Fe/H] abundance variations at this level; here we consider
only those where the [Fe/H] abundance variations are substantially larger.  Second, while the stellar
system Terzan~5 does contain a significant [Fe/H] abundance range, \cite[Massari et al.\ (2015)]{Ma15} argue 
that it is a fossil remnant of the formation of the Galactic bulge, and that it does not have an external accretion origin.  
It will not be considered further here.  Further, while \cite[Simmerer et al.\ (2013)] {JS13} have claimed that the 
globular cluster NGC~3201 contains a significant [Fe/H] range, that result has been questioned.  For 
example, \cite[Mu\~{n}oz et al.\ (2013)]{Mu13} did not find any evidence to support an abundance range in their 
study of the cluster, and \cite[Mucciarelli et al.\ (2015a)]{Mucc15a} have raised concerns about the analysis 
approach used by 
\cite[Simmerer et al.\ (2013)] {JS13}.  This cluster will also not be considered further.  Moreover, in this contribution 
we will also not consider the complexity of the colour-magnitude diagrams now being revealed by {\it HST}
photometry (\cite[e.g., Milone et al.\ 2015a,b]{AM15a,AM15b}), since the complexity does not seem to be restricted 
solely to the clusters showing [Fe/H] abundance variations.  We also note that the light-element abundance
variations, collectively known as the O-Na anti-correlation and which are seemingly ubiquitous in the globular
cluster population, are also found in the clusters with intrinsic [Fe/H] variations.

The most well-known case of a ``globular cluster'' with an internal abundance range is the stellar system
$\omega$~Cen.  This system has a large spread in [Fe/H] among its member stars (\cite[e.g., Johnson \&
Pilachowski 2010]{JP10}) and there is evidence for multiple populations in both the abundance distribution and
the colour-magnitude diagram.  The unusual properties have led to the common 
speculation that $\omega$~Cen is the nuclear remnant of a tidally disrupted dwarf galaxy. 
\cite[Bekki \& Freeman (2003)]{BF03}
have used dynamical model calculations to show that it is feasible to have the nuclear remnant of a disrupted 
dwarf galaxy end up in a tightly bound orbit similar to that of the present-day $\omega$~Cen.  
Additional support for the
scenario lies in the existence in the solar neighbourhood of field stars whose usually high 
[$s$-process/Fe] abundance ratios correspond to those for $\omega$~Cen stars at similar [Fe/H] values.  
Analysis of the kinematics indicates that these stars are very likely to be tidal debris from the $\omega$~Cen
accretion event (\cite[Majewski et al.\ 2012]{SW12}). 

The most straightforward case of a dwarf galaxy nuclear star cluster with an internal abundance range is the 
globular cluster M54, which lies at the centre of the Sgr dwarf.  As shown by,
for example, \cite[Carretta et al.\ (2010)]{Ca10}, this cluster possesses an intrinsic dispersion in [Fe/H]  
among its member stars: $\sigma_{int}$[Fe/H] = 0.18 dex.  The Sgr
dwarf is currently being tidally disrupted, and once the disruption is complete, M54 will become `just another 
globular cluster in the halo'.  Its origin as the nuclear star cluster of a dwarf galaxy will then be much less obvious.

In addition to $\omega$~Cen and M54, there are currently seven other globular clusters with intrinsic [Fe/H] ranges.
These clusters and some of their properties are given in Table~\ref{tab1}.  
Note that the cluster M22 is included in the
list.  \cite[Mucciarelli et al.\ (2015b)]{Mucc15b} have questioned the existence of an [Fe/H] spread in this cluster, as
for NGC~3201.  However, the \cite[Mucciarelli et al.\ (2015b)]{Mucc15b} results do not explain the
range in [$s$-process/Fe] among the stars in this cluster, nor do they offer any explanation for the observed
range in Ca~{\sc{ii}} triplet line strengths among the giants, which points to the presence of an abundance spread
(\cite[Da Costa et al.\ 2009]{GDaC09}).  We now discuss the common characteristics of this set of objects and their
relevance to a potential connection with dwarf galaxy nuclear star clusters.

\begin{table}
\begin{center}
\caption{Globular clusters with intrinsic [Fe/H] spreads}
\label{tab1}
\begin{tabular}{lrrl}
\hline
Cluster & M$_{V}$$^1$ & R$_{gc}$$^1$ & [Fe/H] Spread Reference\\
 & & (kpc) & \\
\hline
NGC 1851 & --8.33 & 16.6 & Carretta et al.\ (2011) \\
NGC 5139 ($\omega$ Cen) & --10.26 & 6.4 & Johnson \& Pilachowski (2010) \\
NGC 5286 & --8.74 & 8.9 & Marino et al.\ (2015) \\
NGC 5824 & --8.85 & 25.9 & Da Costa et al. (2014) \\
NGC 6273 (M19) & --9.13 & 1.7 & Johnson et al.\ (2015) \\
NGC 6656 (M22) & --8.50 & 4.9 & Marino et al.\ (2009) \\
NGC 6715 (M54) &  --9.98 & 18.9 & Carretta et al.\ (2010) \\
NGC 6864 (M75) & --8.57 & 14.7 & Kacharov et al.\ (2013) \\
NGC 7089 (M2) & --9.03 &  10.4 & Yong et al.\ (2014) \\
\hline
\end{tabular}
\end{center}
\vspace{1mm}
$^1$ Values from the on-line version of the Harris (1996) catalogue.
\end{table}


\subsection{Metallicity Distribution Functions}
There are now a number of Milky Way dwarf spheriodal companion galaxies for which individual values of [Fe/H] for
sizeable samples of red giant stars have been determined, enabling the characterisation of the {\it Metallicity
Distribution Function} (MDF) -- the number of stars as a function of [Fe/H].  For example, \cite[Leaman et al.\ (2013)]
{Le13} give MDFs for six dSphs.  One common feature of the dSph MDFs 
(see also \cite[Kirby et al. 2011a,b]{EK11a,EK11b}) is that they rise relatively slowly on the 
metal-poor side of the metallicity peak.  In other words, there is a large range (often more than 1 dex) in [Fe/H]
between the most metal-poor stars and those at the peak of the distribution.  This is in direct contrast to the 
situation in the globular clusters with [Fe/H] ranges -- in those systems with sufficient stars to form the MDF it
always rises
very sharply on the metal-poor side of the peak in the distribution. For example, in the sample of 55 NGC~5286
red giants observed with GIRAFFE by \cite[Marino et al.\ (2015)]{AM15}, there are none with [Fe/H] $<$ --1.90 
but 14 with --1.90 $\leq$ [Fe/H] $<$ --1.80, an extremely rapid rise in the MDF\@.
\cite[Da Costa \& Marino (2011)]{DM11} illustrate the same effect in a comparison of the MDFs for M22 
and $\omega$~Cen. 
Although the extent of the MDFs on the metal-rich side of the peak abundances are different, with $\omega$~Cen 
having a notably longer tail to higher metallicities, both MDFs show steep rises on the metal-poor side of the peak.

{\it A natural interpretation of the MDF difference is that cluster MDFs represent the outcome of rapid enrichment
processes at high star formation rates consistent with high densities at the centre of a dwarf galaxy during the
formation of a nuclear star cluster}, while the dwarf galaxy MDFs represent the result of a more extended star 
formation process over a larger physical scale (e.g., \cite[Kirby et al. 2011a,b]{EK11a,EK11b}). 

\subsection{Outer Density Profiles}

In the original photographic based work of \cite[Grillmair et al. (1995)]{CG95}, and in the new DECam based work of
Kuzma (ANU PhD thesis), the cluster M2 has been shown to be
surrounded by a diffuse halo of stars extending to at least 250pc in
radius.  This is much larger than the nominal `tidal' radius of the cluster, which is $\sim$40pc.  The outer portion 
of the surface density profile is well described by a power-law with a slope of --2.0 $\pm$ 0.1, and in 2-dimensions, 
the outer structure is relatively symmetrical with little indication of any `tidal tails'.  The situation is 
similar to that in NGC~1851 where \cite[Olszewski et al. (2009)]{EO09} discovered that the cluster is also
surrounded by a large diffuse stellar envelope.  The diameter is $\sim$500pc, again much more extended than the 
tidal radius.  Recent work by \cite[Marino et al. (2014)]{AM14} has verified that the NGC~1851 outer diffuse
envelope is 
unambiguously associated with the cluster, and has revealed that it is dominated by stars whose properties 
match the cluster ``1$^{st}$ generation'' (i.e., the stars with lower [Fe/H] and lower [$s$-process/Fe] -- see following 
sub-section).

The same situation occurs in a third globular cluster with an internal [Fe/H] range, NGC~5824.  Here the
analysis of \cite[Grillmair et al. (1995)]{CG95} reveals that the cluster is also surrounded by an extensive diffuse
outer envelope.  The outer density profile is a power law (slope --2.2 $\pm$ 0.1) and cluster stars are detected 
to $r$ $\approx$ 45$^{\prime}$ (see also \cite[Carballo-Bello et al. 2012]{BC12}).  At the distance of NGC~5824 this 
corresponds to $r$ $\approx$ 420pc or a diameter nearly 1kpc in size.  This is significantly larger than the outer
envelopes surrounding M2 and NGC~1851, which are closer to the centre of the Galaxy and thus potentially
more susceptible to tidal stripping.  Indeed the size of the NGC~5824 outer envelope approaches the extent of 
present-day low-luminosity dwarf galaxies.

{\it While it is possible that in each case we are seeing an envelope of escaped cluster stars, it is equally possible 
that the outer envelopes represent remnant populations from a tidally disrupted dwarf galaxy that have remained
bound to the former nuclear star cluster.}

As regards the other clusters in Table \ref{tab1}, M54 is embedded in the Sgr dwarf galaxy, so it could be said
that this cluster is also surrounded by a `large diffuse outer envelope'.  On the other hand, the density 
profiles of $\omega$~Cen and
M22 do not show any significant extra-tidal structure, but given the locations of these clusters relatively close to
the Galactic centre, any outer envelope is likely to have been stripped off.  As for NGC~5296 and M19, there is
little detailed information available on their surface density profiles, but again these clusters are relatively close
to the Galactic centre.  As regards NGC~6864, the surface density profile in 
\cite[Grillmair et al. (1995)]{CG95} hints at the presence of an outer envelope; a modern study based on 
digital wide-field imaging is required for confirmation.

\subsection{([$s$-process/Fe], [Fe/H]) correlations}

Perhaps the most intriguing characteristic in the clusters with [Fe/H] ranges is the existence of correlation 
between [Fe/H] and [$s$-process element/Fe] abundance ratios: in all clusters where sufficient data exist (seven of
the nine objects), the stars with larger [Fe/H] abundances also have higher [$s$-element/Fe] abundance
ratios\footnote{In the case of $\omega$~Cen, which has the largest [Fe/H] range, the effect appears to saturate in the 
sense that the [$s$-process/Fe] ratios reach a plateau and do not continue to increase with increasing [Fe/H]
beyond [Fe/H] $\approx$ --1.3.}.
The correlation is unexpected in a nucleosynthetic sense because it requires an additional $s$-process 
element contribution, most probably from AGB-stars, over and above the contribution needed to maintain the 
abundance ratio as the iron abundance increases.  An example of the correlation for the cluster NGC~5286 is
shown in Fig.\ 6 of \cite[Marino et al. (2015)]{AM15}, who point out two further
pieces of information.  First, the relative enhancement of different neutron-capture elements correlates with the 
$s$-process element fraction in material with the solar composition.  This verifies that the enrichment does indeed 
involve $s$-process nucleosynthesis.  Second, a comparison of a subset of the clusters (NGC~5286, M2, M22,
and $\omega$~Cen -- see Fig.\ 19 of  \cite[Marino et al. 2015]{AM15}) shows that the rate of increase in 
\mbox{[$s$-process/Fe]} with [Fe/H] is similar but not identical among these clusters.  NGC~5286 has the steepest
gradient while that for M22 and $\omega$~Cen are somewhat shallower.  The nucleosynthesis
process is therefore apparently similar from cluster-to-cluster, but not identical.    It is, however, not by any 
means clear how this correlation fits into the ``[Fe/H abundance range implies a former nuclear star cluster'' 
hypothesis.

\section{Discussion}

The information presented above for the nine clusters with [Fe/H] spreads can be summarised as 
follows. (i) M54 is the nuclear star cluster of the Sgr dwarf.  (ii) $\omega$~Cen is almost certainly the nuclear 
remnant of a dwarf galaxy that has been accreted and disrupted by the Milky Way.  (iii) The distribution of [Fe/H] 
values in the clusters consistently shows steep rises on the metal-poor side, which is consistent with rapid 
enrichment at a central location. (iv) At least some of the clusters in question are surrounded by extended 
stellar envelopes that might represent the remnant population of an accreted dwarf galaxy. 

It is then important to note that \cite[Georgiev et al. (2009)] {Ge09} have shown that the nuclear star clusters in 
current dwarf galaxies have similar properties to those for luminous Milky Way globular clusters as regards 
luminosity (M$_{V}$), and size (half-light radius r$_{h}$).  
{\it These results then make it plausible that the clusters with internal [Fe/H] ranges are the former nuclear star
clusters of dwarf galaxies accreted and disrupted by the Milky Way.}
However, this inference can only be valid if the number of such former nuclear star clusters is consistent with
other approaches to the issue.  For example, the current number of clusters with [Fe/H] ranges is broadly
consistent with the results of \cite[Pffeffer et al. (2014)]{Pf14}, who have used cosmological simulations to
suggest 1--3 massive Milky Way globular clusters are former dwarf galaxy nuclei.  Another estimate of the
expected number of former dwarf galaxy nuclear star clusters can be made as follows.  The absolute visual
magnitude of the Galactic stellar halo is approximately M$_{V}$ $\approx$ --17 \cite[(e.g., Freeman 1993)]{KF93}.  If
we assume that $\sim$50\% of this luminosity comes from the disruption of satellite galaxies with the rest
formed in-situ, and if we further assume that the disrupted systems had comparable luminosites to the present-day
Fornax or Sgr systems, i.e., M$_{V}$ $\approx$ --14, then the disruption of $\sim$15 systems can provide the
required luminosity.  If we then further assume that $\sim$50\% of the disrupted dwarfs had nuclear star
clusters, then we arrive at the, admittedly uncertain, estimate of approximately eight such systems 
in the Milky Way halo.  As
noted in Table \ref{tab1} there are nine known ``globular clusters'' with internal [Fe/H] spreads.  The numbers are
thus consistent but clearly if substantially more globular clusters are found to possess significant internal [Fe/H]
abundance spreads, the hypothesized connection between such clusters and the nuclear star clusters of disrupted
dwarfs will require re-consideration.

\begin{figure}
\begin{center}
\includegraphics[width=0.55\textwidth,angle=-90.]{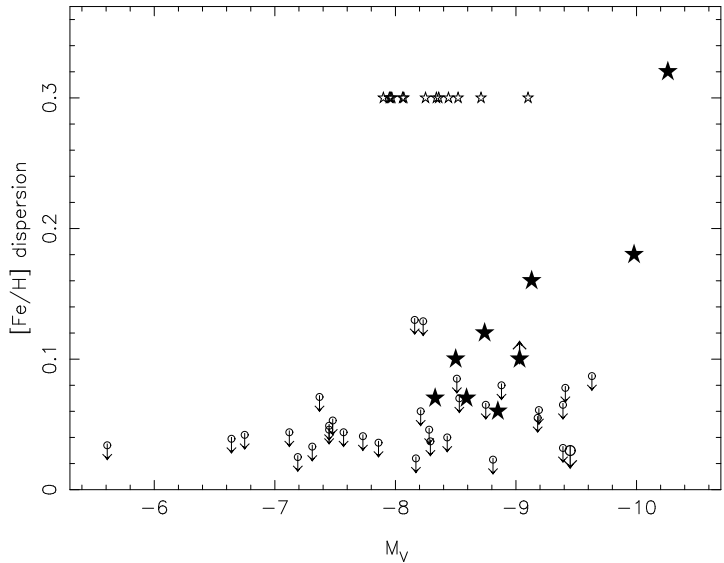} 
 \caption{Metallicity dispersion, or upper limit, for Milky Way globular clusters plotted against absolute visual
 magnitude.  The sample of clusters included is complete for M$_{V} < -7.9$, fainter clusters are taken from
 Carretta et al.\  (2010).  The filled star symbols are the 9 clusters with intrinsic [Fe/H] dispersions.  The 13 unstudied
 (or poorly studied) clusters with M$_{V} < -7.9$ are shown in the upper part of the plot as open star symbols.}
   \label{fig2}
\end{center}
\end{figure}

To assess this question we show in Fig.\ \ref{fig2} an updated version of a plot first presented by 
\cite[Carretta et al. (2010)]{Ca10}.  In this plot we show either an upper limit on the potential [Fe/H] abundance 
range present, or abundance dispersion estimates for the clusters in Table \ref{tab1}.  The literature has been 
searched to include upper limits for all Milky Way clusters with M$_{V} < -7.9$; for the fainter clusters the limits
shown are only
for the clusters studied by \cite[Carretta et al. (2010)]{Ca10}.  As indicated in the figure, there are 13 
unstudied or poorly studied relatively luminous (M$_{V}$ $<$ --7.9) Milky 
Way globular clusters.   Not surprisingly, these clusters are mostly at large distances from the Sun and/or have large
reddenings.  Five have E$(B-V)$ $<$ 0.3 mag: NGC~5024, 6541, 5986, 6229 and 6284.  In the
terminology of \cite[Lee et al. (2007)]{YWL07} NGC~5986 has a strongly extended blue HB (like $\omega$~Cen,
M22, M54 and M2) and NGC~6629 has a moderately extended blue HB (like NGC~1851, 5824 and 6864).
Detailed studies of these clusters would be very worthwhile.

In summary, the hypothesis that the globular clusters with substantial internal [Fe/H] abundance ranges are the
former nuclear star clusters of now disrupted dwarf galaxies has to be considered at least plausible.  However, 
further work is required is to substantiate the total number of such clusters -- if many more are discovered then the  
hyphothesis would need revision.

\end{document}